\documentclass{article}

\usepackage[dblblindworkshop, final, nonatbib]{neurips_2026}
\workshoptitle{The 9th Annual Workshop for Machine Learning \& the Physical Sciences (ML4PS 2026)}
\makeatletter
\renewcommand{\@trackname}{\@workshoptitle.}
\makeatother
\usepackage{algorithm,algpseudocode}

\usepackage{amsthm}

\usepackage{amssymb}

\usepackage{enumitem}
\usepackage[utf8]{inputenc}%
\usepackage[T1]{fontenc}%
\usepackage{hyperref}%
\usepackage{url}%
\usepackage{booktabs}%
\usepackage{amsfonts}%
\usepackage{amsmath}
\usepackage{nicefrac}%
\usepackage{microtype}%
\usepackage{xcolor}%
\usepackage{cleveref}
\usepackage{graphicx}
\usepackage{multicol,multirow}
\usepackage{xcolor}
\usepackage{makecell}
\usepackage{graphicx}
\usepackage{booktabs}
\usepackage{caption}
\usepackage{aas_macros}

\newcommand{\vs}[2]{\shortstack{#1\\[0.5pt]{\tiny$\pm$#2}}}
\newcommand{\vsb}[2]{\shortstack{\textbf{#1}\\[0.5pt]{\tiny\textbf{$\pm$#2}}}}
\newcommand{\vsu}[2]{\shortstack{\underline{#1}\\[0.5pt]{\tiny$\pm$#2}}}

\title{Probabilistic Inversion of Strong Gravitational Lenses with Flow Matching Priors}

\author{%
  \textbf{Tianao Li$^{1,5}$, Jimena González$^{1,5}$, Xinfeng Xu$^{1,3,5}$, Aggelos K. Katsaggelos$^{1,5}$,}\\
  \textbf{Rebecca Willett$^{2,5}$, Aleksandra \'Ciprijanovi\'c$^{2,4,5}$, Tjitske Starkenburg$^{1,5}$, Emma Alexander$^{1,5}$} \\
  $^1$Northwestern University, $^2$University of Chicago, $^3$Zhejiang University, \\
  $^4$Fermi National Accelerator Laboratory, $^5$NSF-Simons AI Institute for the Sky (SkAI) \\
  \url{https://nubivlab.github.io/Blind-FlowSGS/} \\
}

\begin{document}

\maketitle

\begin{abstract}
  Strong gravitational lensing reveals important information about the universe, but reconstructing the source image and lens properties requires solving an ill-posed inverse imaging problem.
  Generative image models serve as powerful data-driven priors in Bayesian inverse problems.
  Diffusion models have been used widely, but they are computationally costly.
  Flow matching has emerged as a promising alternative to diffusion models, with easier training, efficient sampling, and better-quality data generation.
  In this work, we present {\bf Blind-FlowSGS}, a principled blind posterior sampling method using flow matching priors, enabling efficient posterior sampling for non-blind and blind inversion of strong gravitational lenses.
  On simulated Euclid data, we demonstrate our method's improvement over diffusion-based methods in both efficacy and efficiency.
  Our method shows promising capabilities to accurately process strong lensing data from modern sky surveys at scale.
  Our code and data are available on our project page.
\end{abstract}

\vspace{-1.7mm}
\section{Introduction}
\label{sec:intro}
\vspace{-1.7mm}

Strong gravitational lensing is the phenomenon that occurs when the light of an astronomical source is bent by the gravitational potential of a massive foreground object, producing distorted images of the source~\cite{treu2010strong,Shajib_2024_slreview,Einstein1936LENSLIKEAO}. Strong lenses have multiple science applications: individual systems can be used to measure the Hubble constant~\cite{Treu_2022, Shajib_2020} and probe the small-scale distribution of dark matter~\cite{vegetti2023_sl_dm}, while large lens populations can constrain the nature of dark energy~\cite{Li_2024_largepopulation}. In order to use strong lenses in scientific applications, it is necessary to model their properties accurately. Traditionally, strong lens modeling involves parameterizing the mass and light distribution of the objects, and using density samplers to find and optimize those parameters by comparing a simulation with the real observation~\cite{brewer2006strong}. However, this computationally expensive approach is not feasible to model the $\sim10^5$ lenses that next-generation astronomical surveys like LSST and Euclid will uncover~\cite{Collett_2015, Shajib_2023_SLfromVeraRubin, Euclid_engineA}. Recent machine learning (ML) approaches to lens parameter prediction include training ML models in a regression task for direct inference~\cite{Hezaveh_2017, Schuldt_2023, Pearson_2021}, with the LEMON pipeline~\cite{euclid2026lemon} demonstrating this at scale on real Euclid Q1 lenses~\cite{Euclid_engineA}, and simulation-based inference (SBI), which trains on simulations to yield well-calibrated posteriors without an explicit likelihood~\cite{Poh_2025, Legin_2021, Erickson_2025}.

With the rapid advancements in generative modeling, diffusion models~\cite{song2019generative,ho2020denoising,song2021scorebased} have been used as priors to solve Bayesian inverse problems~\cite{chung2023diffusion,wu2024principled,zhang2025improving,murata2023gibbsddrm,sanghvi2024kernel,hu2026prism}, including reconstruction of strong gravitational lenses.
Specifically,~\cite{adam2022posterior,karchev2022strong} used diffusion models to reconstruct the source galaxy assuming a known lens, and~\cite{barco2025blind} used diffusion models for joint inference of the source and lens.
Among these,~\cite{barco2025blind} is the closest prior work to ours: it jointly infers the source and a parametric lens mass profile with a score-based prior, using a continuous-time sampler derived from GibbsDDRM~\cite{murata2023gibbsddrm}, and reports residuals consistent with the observational noise and lens-parameter marginals without systematic bias. We instead use a flow matching prior with split Gibbs sampling~\cite{vono2019split} in the source block, and informative lens priors estimated from Euclid Q1.
Recently, flow matching~\cite{lipman2023flow,liu2023flow,albergo2023building} has emerged as the next-generation generative model that improves upon diffusion.
Flow matching constructs straighter probability paths, enabling easier training, more efficient sampling, and better data-generation quality.
Flow matching has been used to solve classic (non-blind, linear) inverse problems~\cite{martin2025pnpflow,patel2025flowchef,kim2025flowdps}.
It has also been applied to strong lensing in other roles, namely amortized posterior inference of lens parameters with simulator feedback~\cite{holzschuh2024flowmatching} and fast emulation of lensed images~\cite{sayed2025flowlensing}, but not as a plug-and-play image prior for pixel-level source reconstruction, and not for blind inference of the source and the lens together. This is the gap we address, and it matters for efficiently processing the large data volumes expected from modern wide-area astronomical surveys.

In this work, we focus on inversion of galaxy-galaxy lenses and parameterize the light distribution of the source galaxy as a pixelated image $\mathbf{x}$ and the mass distribution of the deflector as a singular isothermal ellipsoid (SIE) profile~\cite{GravitationalLensing} with parameters $\boldsymbol{\psi}\in\mathbb{R}^{5}$ (see~\Cref{tab:priors}). We consider both the blind and non-blind settings. In the blind setting, given an observed image $\mathbf{y}$, we aim to sample from the joint posterior distribution of the source and lens parameters
\begin{equation}
    \mathbf{x}, \boldsymbol{\psi} \sim
    p(\mathbf{x}, \boldsymbol{\psi}\vert\mathbf{y})
    \propto p(\mathbf{y}\vert\mathbf{x}, \boldsymbol{\psi})p(\mathbf{x})p(\boldsymbol{\psi}),
    \label{eq:blind}
\end{equation}
where $p(\mathbf{y}\vert\mathbf{x}, \boldsymbol{\psi}) = \mathcal{N}(\mathbf{y};\mathcal{A}(\mathbf{x}; \boldsymbol{\psi}),\sigma^2\mathbf{I})$ is the data likelihood assuming Gaussian noise with noise level $\sigma$, and $\mathcal{A}(\cdot)$ is the lensing forward model. Assuming a known lens $\boldsymbol{\psi}$, this blind inverse problem reduces to a non-blind inverse problem:
\begin{equation}
    \mathbf{x} \sim
    p(\mathbf{x}\vert\mathbf{y}, \boldsymbol{\psi})
    \propto p(\mathbf{y}\vert\mathbf{x}, \boldsymbol{\psi})p(\mathbf{x}).
    \label{eq:nonblind}
\end{equation}

To sample from the posterior in Eq.~\eqref{eq:nonblind}, we leverage a new principled posterior sampler~\cite{li2026flowsgs} using flow matching as a prior, and extend it to the blind scenario in Eq.~\eqref{eq:blind}.
We train flow matching models on galaxy images from the IllustrisTNG simulations~\cite{nelson2019first, pillepich2019first, nelson2021illustristngsimulationspublicdata, baes2024tngskirtatlas} for the source prior $p(\mathbf{x})$ and test our method on simulated strong lensing systems using Euclid's parameters.
We also incorporate non-uniform priors for the deflector mass parameters $\boldsymbol{\psi}$ estimated from $\sim$330 real strong lenses discovered in the Euclid Q1 data release~\cite{Euclid_engineA}.
In both source-only reconstruction and joint source-lens reconstruction experiments, we benchmark a variety of diffusion-based and flow-based methods and demonstrate our method's efficacy and efficiency.

\begin{figure*}[!t]
    \centering
    \includegraphics[width=1.0\linewidth]{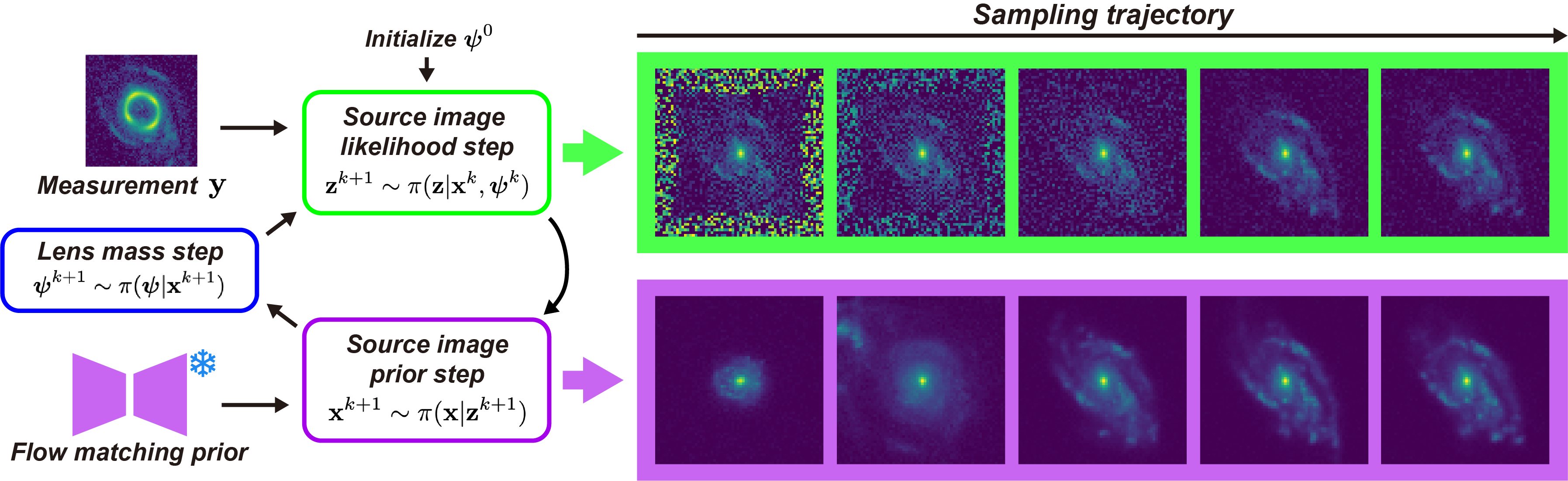}
    \caption{
        \textbf{A high-level overview of Blind-FlowSGS, a principled blind posterior sampler using flow matching as a plug-and-play prior.}
        Our method iterates between a {\color{green} source image likelihood step}, a {\color{violet} source image prior step}, and a {\color{blue} lens mass step}.
        Here $\mathbf{z}$ is the auxiliary variable of the source image split.
        We show sampling trajectories of the source galaxy (right).
        See pseudocode of Blind-FlowSGS in~\Cref{alg:blind-flowsgs}.
        Adapted from~\cite{li2026flowsgs}.
    }
    \label{fig:teaser}
\end{figure*}

\vspace{-1.8mm}
\section{Methods}
\label{sec:methods}

\begin{algorithm}[t]
    \caption{Blind (joint source and lens parameter) posterior sampling with Blind-FlowSGS.}
    \label{alg:blind-flowsgs}
    \begin{algorithmic}[1]
        \Require measurement $\mathbf{y}$, forward model $\mathcal{A}(\cdot)$, lens parameter prior $p(\boldsymbol{\psi})$, whitening map $\mathcal{T}:\boldsymbol{\psi}\mapsto\mathbf{u}$ with Jacobian $\mathbf{J}(\mathbf{u})=\partial\boldsymbol{\psi}/\partial\mathbf{u}$, measurement noise level $\sigma$, velocity network $\mathbf{v}_\theta(\cdot,\cdot)$ pretrained with interpolant schedule $(\alpha_t,\sigma_t)$, diffusion coefficient $w_t$, number of iterations $K$, coupling parameters $\{\rho_k\}_{k=0}^{K-1}$, Langevin step size $\eta$ and inner steps $J$
        \State $\mathbf{x}^0 \sim \mathcal{N}(0,\mathbf{I})$, \quad $\mathbf{u}^0 \sim p(\mathbf{u})$, \quad $\boldsymbol{\psi}^0 \leftarrow \mathcal{T}^{-1}(\mathbf{u}^0)$ \Comment{Initialization.}
        \For {$k = 0, 1, \ldots, K-1$}
            \State $\mathbf{z}^{k+1} \sim \pi(\mathbf{z} \vert \mathbf{x}^{k},\boldsymbol{\psi}^{k}) \propto \exp\left[-\frac{\|\mathcal{A}(\mathbf{z};\boldsymbol{\psi}^{k}) - \mathbf{y}\|^2}{2\sigma^2} - \frac{\|\mathbf{x}^{k} - \mathbf{z}\|^2}{2\rho_k^2}\right]$ \Comment{\textcolor{green}{\bf Source likelihood step.}}
            \State \hspace{\algorithmicindent} $\mathcal{A}(\cdot\,;\boldsymbol{\psi}^{k})$ is affine in $\mathbf{z}$, so this Gaussian is sampled exactly by randomize-then-optimize.
            \State $t_k \leftarrow \left(\frac{\sigma_t}{\alpha_t}\right)^{-1}(\rho_k)$ \Comment{\textcolor{violet}{\bf Compute time $t_k$ corresponding to noise level $\rho_k$.}}
            \State $\mathbf{x}_{t_k} \leftarrow \alpha_{t_k}\, \mathbf{z}^{k+1}$ \Comment{\textcolor{violet}{\bf Scale $\mathbf{z}^{k+1}$ to match noise level at $t_k$.}}
            \State $\mathbf{x}^{k+1} \leftarrow \text{Euler–Maruyama}(\text{SDE},\mathbf{x}_{t_k},t_k,0)$ \Comment{\textcolor{violet}{\bf Source prior step.}}
            \State \hspace{\algorithmicindent} where the SDE is defined by
            $\mathrm{d}\mathbf{x}_t
            = \left[\left(1 - \frac{w_t \alpha_t}{2\sigma_t \gamma_t}\right)\mathbf{v}_\theta(\mathbf{x}_t, t) + \frac{w_t \dot\alpha_t}{2\sigma_t \gamma_t}\,\mathbf{x}_t\right]\mathrm{d}t + \sqrt{w_t}\,\mathrm{d}\bar{\mathbf{w}}_t,$
            \State \hspace{\algorithmicindent} and $\gamma_t = \dot\alpha_t\sigma_t - \alpha_t\dot\sigma_t$.
            \State $\mathbf{w}^{(0)} \leftarrow \mathbf{u}^{k}$
            \For {$j = 0,1,\ldots,J-1$} \Comment{\textcolor{blue}{\bf Lens mass step.}}
                \State $\mathbf{g} \leftarrow \mathbf{J}(\mathbf{w}^{(j)})^\top \nabla_{\boldsymbol{\psi}}\frac{\|\mathcal{A}(\mathbf{x}^{k+1};\boldsymbol{\psi}) - \mathbf{y}\|^2}{2\sigma^2}\Big|_{\boldsymbol{\psi}=\mathcal{T}^{-1}(\mathbf{w}^{(j)})} - \nabla_{\mathbf{u}} \log p(\mathbf{u})\big|_{\mathbf{u}=\mathbf{w}^{(j)}}$
                \State $\mathbf{w}^{(j+1)} \leftarrow \Pi\!\left[\mathbf{w}^{(j)} - \eta\,\mathbf{g} + \sqrt{2\eta}\,\boldsymbol{\xi}^{(j)}\right],\quad \boldsymbol{\xi}^{(j)}\sim\mathcal{N}(0,\mathbf{I})$
            \EndFor
            \State $\mathbf{u}^{k+1} \leftarrow \mathbf{w}^{(J)}$, \quad $\boldsymbol{\psi}^{k+1} \leftarrow \mathcal{T}^{-1}(\mathbf{u}^{k+1})$
        \EndFor
        \State \Return $\mathbf{x}^K,\ \boldsymbol{\psi}^K$
    \end{algorithmic}
\end{algorithm}

\begin{figure*}[b]
    \centering
    \includegraphics[width=1.0\linewidth]{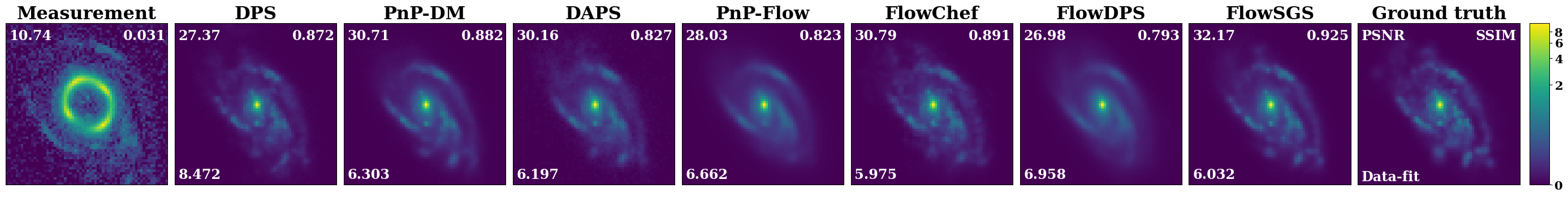}
    \caption{
        \textbf{Results on source-only (non-blind) reconstruction.}
        FlowSGS provides better image quality (PSNR, SSIM) than previous posterior samplers.
    }
    \label{fig:nonblind}
\end{figure*}

\vspace{-1.7mm}
\subsection{Non-blind posterior sampling}
\label{sec:flowsgs}
\vspace{-1.7mm}

We apply the FlowSGS sampler~\cite{li2026flowsgs} to sample from the posterior in Eq.~\eqref{eq:nonblind} for the strong gravitational lensing problem. FlowSGS follows the plug-and-play~\cite{venkatakrishnan2013plug,chan2016plug} split Gibbs sampling (SGS) framework~\cite{vono2019split}, extending the diffusion-based PnP-DM~\cite{wu2024principled} to flow matching. 
FlowSGS alternates between a likelihood step and a prior step, sharing a similar ``divide-and-conquer'' spirit with HQS~\cite{charbonnier1994two,geman1995nonlinear} and ADMM~\cite{boyd2004convex,neal2011distributed} (see~\Cref{alg:flowsgs}).
SGS avoids most approximations used in likelihood-gradient guidance or gradient-free vector-field steering~\cite{chung2023diffusion,martin2025pnpflow,patel2025flowchef,kim2025flowdps}, so it offers a principled way to incorporate measurement likelihood and produce a more accurate posterior (see~\Cref{tab:nonblind} and~\Cref{fig:nonblind}). This method relies on a generative prior for source galaxy images; see~\Cref{fig:samples} for samples from our flow prior and~\Cref{sec:sim} for training details to ensure a high-quality prior.

\vspace{-1.5mm}
\subsection{Blind posterior sampling}
\label{sec:Blind-FlowSGS}
\vspace{-1.5mm}

Following PRISM~\cite{hu2026prism}, we extend FlowSGS into a blind posterior solver, which we call \textbf{Blind-FlowSGS} (see~\Cref{fig:teaser}). Specifically, we target the joint posterior in Eq.~\eqref{eq:blind} with a Gibbs sweep that alternates between sampling the source $\mathbf{x}$ and the parameters $\boldsymbol{\psi}$, applying split Gibbs sampling in the source block. With $\boldsymbol{\psi}$ fixed, the source block is exactly the non-blind sampler of~\Cref{sec:flowsgs}: a likelihood step followed by a prior step that integrates the reverse-time SDE with the pretrained flow matching prior. With $\mathbf{x}$ fixed, the parameter block is a projected Langevin chain targeting $p(\boldsymbol{\psi}\mid\mathbf{x},\mathbf{y}) \propto p(\mathbf{y}\mid\mathbf{x},\boldsymbol{\psi})p(\boldsymbol{\psi})$ under the Euclid Q1 priors (see~\Cref{tab:priors} and~\Cref{fig:prior}). We sample $\boldsymbol{\psi}$ in whitened coordinates, in which every parameter carries unit prior scale, so that a single step size serves all 5 parameters despite their prior scales differing by about an order of magnitude. We define the whitening map $\mathcal{T}$ and the projection $\Pi$ in~\Cref{app:prior}.
The pseudocode of FlowSGS is shown in~\Cref{alg:flowsgs}, with our novel extension to the blind setting in Algorithm~\ref{alg:blind-flowsgs}.
We also use Meta Flow Maps (MFM)~\cite{potaptchik2026meta} to accelerate both FlowSGS and Blind-FlowSGS, referred to as ``FlowSGS (MFM)'' and ``Blind-FlowSGS (MFM)''.
See~\cite{wu2024principled,li2026flowsgs,hu2026prism} for details on the SGS variable splitting, mathematical approximations, convergence guarantees, and derivations.

\vspace{-1.7mm}
\section{Experiments}
\label{sec:exp}

\begin{figure}[b]
\centering
\begin{minipage}[t]{0.52\textwidth}
  \vspace{0pt}%
  \centering
  \captionof{table}{
      \textbf{Quantitative evaluation on source-only (non-blind) reconstruction.} 
      All methods draw 8 posterior samples, and the reported metrics are computed on the posterior mean source galaxy image, with standard deviations across the 100 test lenses.
    }
  \label{tab:nonblind}
  \footnotesize
  \setlength{\tabcolsep}{1.5pt} 
  \begin{tabular}{lcccc}
    \toprule
    \bf Method & \bf PSNR ($\uparrow$) & \bf SSIM ($\uparrow$) & \bf NFE & \bf Time (s) \\
    \midrule
    DPS~\cite{chung2023diffusion} & 29.06{\tiny$\pm$1.16} & 0.885{\tiny$\pm$0.013} & $1000$ & $23.8$ \\
    PnP-DM~\cite{wu2024principled} & 34.64{\tiny$\pm$0.65} & 0.925{\tiny$\pm$0.007} & $1619$ & $16.1$ \\
    DAPS~\cite{zhang2025improving} & 33.15{\tiny$\pm$0.56} & 0.851{\tiny$\pm$0.005} & $1000$ & $122.1$ \\
    PnP-Flow~\cite{martin2025pnpflow} & 26.01{\tiny$\pm$2.27} & 0.776{\tiny$\pm$0.025} & $100$ & $1.7$ \\
    FlowChef~\cite{patel2025flowchef} & 33.74{\tiny$\pm$0.64} & 0.900{\tiny$\pm$0.007} & $100$ & $1.7$ \\
    FlowDPS~\cite{kim2025flowdps} & 31.15{\tiny$\pm$0.64} & 0.877{\tiny$\pm$0.012} & $100$ & $2.9$ \\
    \bf FlowSGS~\cite{li2026flowsgs} & \underline{35.19{\tiny$\pm$0.66}} & \underline{0.929{\tiny$\pm$0.006}} & $881$ & $9.7$ \\
    \bf FlowSGS (MFM)~\cite{li2026flowsgs} & \textbf{35.43{\tiny$\pm$0.67}} & \textbf{0.932{\tiny$\pm$0.006}} & $212$ & $3.9$ \\
    \bottomrule
  \end{tabular}
\end{minipage}\hfill
\begin{minipage}[t]{0.45\textwidth}
  \vspace{0pt}
  \centering
  \includegraphics[width=\linewidth]{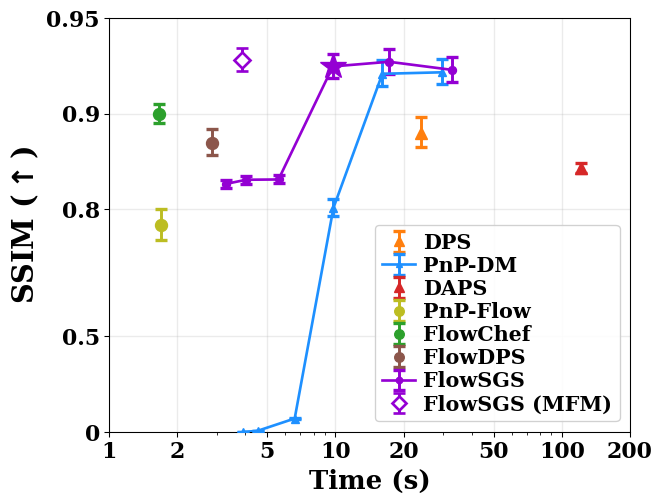}
  \caption{
      \textbf{Time-performance trade-offs on source-only (non-blind) reconstruction.}
  }
  \label{fig:steps}
\end{minipage}
\end{figure}

\vspace{-1.6mm}
\subsection{Dataset and generative models}
\label{sec:sim}
\vspace{-1.5mm}

To train our flow matching prior, as well as a diffusion prior for baseline comparisons, we generate simulated data matching Euclid's observational parameters. Our simulated data enables access to noise-free ground truth and is based on synthetic images of galaxies from the IllustrisTNG TNG50-1 run~\cite{nelson2019first,pillepich2019first,nelson2021illustristngsimulationspublicdata,baes2024tngskirtatlas}, which are resampled to the pixel scale of Euclid's VIS instrument (0.1 arcsec) at 64$\times$64 pixels.
Among these processed galaxy images, we randomly select 10,000 to train the generative priors and 100 to simulate strong gravitational lens systems. 
Importantly, we find that normalized flux images (99th percentile pixel set to 1.0) stabilize generative model training, compared to physical values. 
We also simulate measurements assuming normalized source images.
We train flow and diffusion models using the NCSN++ architecture~\cite{song2021scorebased} for fair comparisons, and use the Variance Preserving (VP)~\cite{song2019generative,song2021scorebased} SDE as the diffusion schedule and the linear interpolant~\cite{liu2023flow} for flow matching. 
Both models are trained for 1,000 epochs using the Adam optimizer with a learning rate of $1\times10^{-4}$.
We show several unconditional samples from the pretrained flow model in~\Cref{fig:samples}. We estimate the distributions of the lens parameters $\boldsymbol{\psi}$ from real strong lenses in the Euclid Q1 data release~\cite{Euclid_engineA} (see~\Cref{tab:priors}), which we use for data generation and the lens prior in Eq.~\eqref{eq:blind}.
We use Caustics~\cite{stone2024caustics} to simulate 64$\times$64 lensed images and add i.i.d. Gaussian noise with $\sigma=0.1$ to obtain the observed images; we do not include the telescope PSF.

\begin{figure}[!t]
\centering

\captionof{table}{
    \textbf{Quantitative evaluation on joint source and lens (blind) reconstruction.}
    All results are computed across 128 source samples and 1,024 lens samples, with standard deviations across the 100 test lenses. The mean absolute error (MAE) is per lens mass parameter, in arcsec for $x_0$, $y_0$, $\theta_{\rm E}$ and in rad for $\phi$.
    $^{\dagger}$Note that GibbsDDRM~\cite{murata2023gibbsddrm,barco2025blind} also backpropagates through the denoiser at every evaluation, adding 400 backward passes.
}
\label{tab:blind}
\footnotesize
\setlength{\tabcolsep}{2pt}
\begin{tabular}{lccccccccc}
\toprule
\multirow{2}{*}{\bf Method} & \multirow{2}{*}{\bf PSNR ($\uparrow$)} & \multirow{2}{*}{\bf SSIM ($\uparrow$)} & \multicolumn{5}{c}{\bf MAE ($\downarrow$)} & \multirow{2}{*}{\bf NFE} & \multirow{2}{*}{\bf Time (s)} \\
\cmidrule(lr){4-8}
& & & $x_0$ & $y_0$ & $q$ & $\phi$ & $\theta_{\rm E}$ & & \\
\midrule
GibbsDDRM~\cite{murata2023gibbsddrm,barco2025blind} & \vs{28.59}{0.75} & \vs{0.874}{0.011} & \vs{0.0178}{0.0031} & \vs{0.0173}{0.0028} & \vs{0.033}{0.004} & \vs{0.127}{0.025} & \vs{0.0475}{0.0123} & $400 + 400^{\dagger}$ & \underline{146} \\
PRISM~\cite{hu2026prism} & \vsu{30.36}{0.78} & \vsu{0.887}{0.012} & \vsu{0.0153}{0.0027} & \vs{0.0141}{0.0026} & \vs{0.024}{0.004} & \vsb{0.078}{0.017} & \vsu{0.0288}{0.0095} & 2739 & 285 \\
\bf Blind-FlowSGS & \vsu{30.36}{0.85} & \vsu{0.887}{0.013} & \vsb{0.0149}{0.0028} & \vsb{0.0130}{0.0025} & \vsb{0.021}{0.003} & \vsu{0.087}{0.018} & \vs{0.0299}{0.0109} & 1628 & 219 \\
\bf Blind-FlowSGS (MFM) & \vsb{31.21}{0.78} & \vsb{0.920}{0.009} & \vsu{0.0153}{0.0029} & \vsu{0.0136}{0.0024} & \vsu{0.022}{0.003} & \vs{0.090}{0.019} & \vsb{0.0287}{0.0148} & \textbf{211} & \textbf{133} \\
\bottomrule
\end{tabular}

\end{figure}

\begin{figure*}[t]
    \centering
    \includegraphics[width=0.98\linewidth]{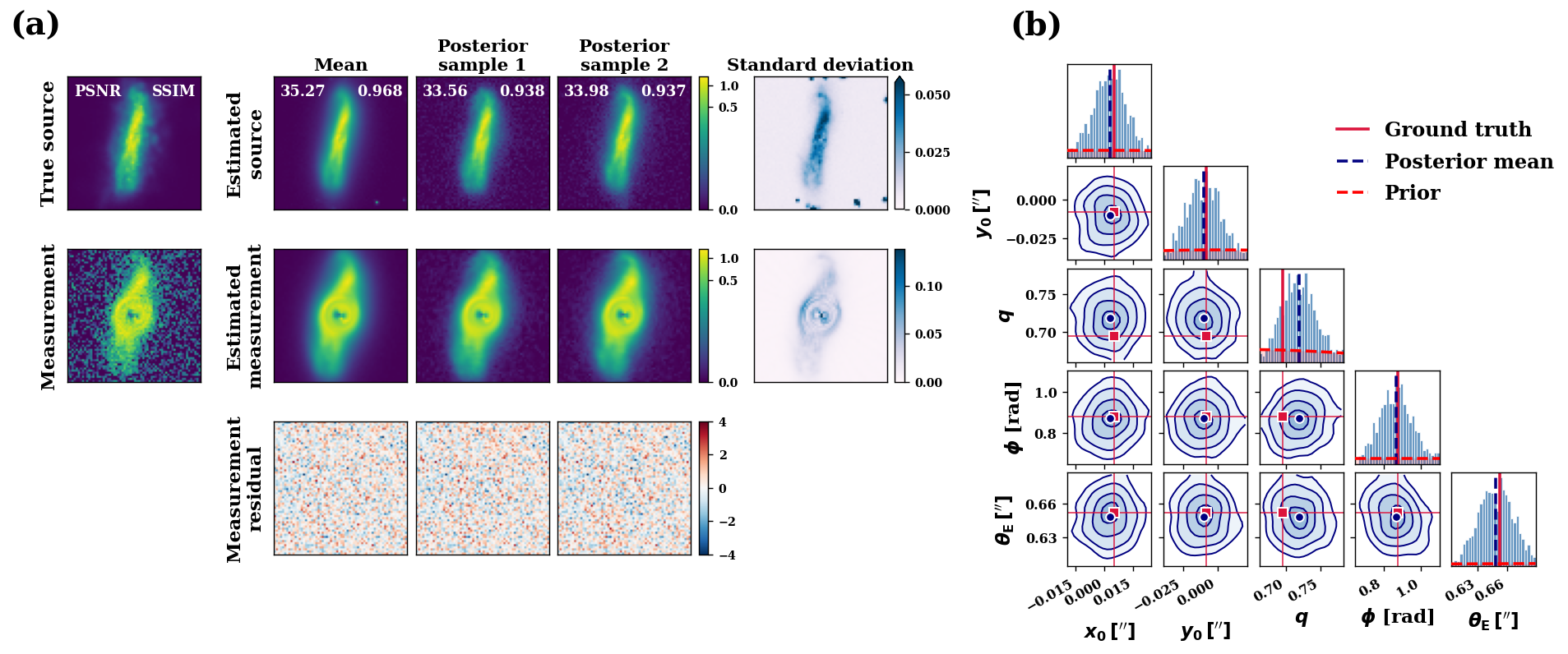}
    \caption{
        \textbf{Results on joint source and lens (blind) reconstruction.}
        (a) Source galaxy image, measurement, and measurement residuals, shown in mean, posterior samples, and standard deviations.
        (b) Corner plots for 5 SIE lens mass parameters computed from 1,024 posterior samples.
        Red dashed lines on the 1D marginals show the lens parameter priors of~\Cref{tab:priors}.
    }
    \label{fig:blind}
\end{figure*}

\vspace{-1.5mm}
\subsection{Source-only (non-blind) reconstruction}
\label{sec:nonblind}
\vspace{-1.5mm}

We test FlowSGS on 100 simulated strong gravitational lenses whose source images are not used during training and report the image quality metrics in~\Cref{tab:nonblind}.~\Cref{fig:nonblind} shows a visual comparison between all considered methods, where FlowSGS provides the best image quality, followed by PnP-DM~\cite{wu2024principled}, a principled diffusion posterior sampler using a similar SGS framework.
To further demonstrate efficiency, we compare FlowSGS (purple) with PnP-DM (blue) and vary the number of SDE discretization steps in the prior step, shown in~\Cref{fig:steps}.
We report the runtime of drawing a batch of 8 posterior samples on a single NVIDIA RTX A6000 GPU and the number of function evaluations (NFE).
FlowSGS outperforms all other methods and is much more robust to few-step sampling in the prior step, enabling efficient sampling with minimal sacrifice in accuracy.
By default, FlowSGS uses only 32 SDE discretization steps (marked with a purple star) while PnP-DM requires 100.
This demonstrates a significant efficiency advantage thanks to flow matching's straighter probability paths and is beneficial for processing large amounts of data from sky surveys.

\vspace{-1.5mm}
\subsection{Joint reconstruction of source and lens (blind)}
\label{sec:blind}
\vspace{-1.5mm}

We apply our Blind-FlowSGS sampler to jointly reconstruct the source and lens. 
In~\Cref{fig:blind}a, we show the mean and standard deviations of the source galaxy image, the estimated observations, and the observation residuals.
Because sampling a large number of source images is computationally expensive, we follow~\cite{barco2025blind} and draw 128 joint source and lens samples, and then use the unadjusted Langevin algorithm (ULA)~\cite{roberts1996exponential} to sample 7 more lens samples per source sample from the conditional $p(\boldsymbol{\psi}\vert\mathbf{x},\mathbf{y})$.
This gives us 128 source samples and 1,024 lens samples in total. We use the same procedure for all blind samplers in~\Cref{tab:blind}.
In~\Cref{fig:blind}b, we show the corner plots of the 5 lens mass parameters (see results on more lensing systems in Appendix~\ref{app:results}).
We benchmark Blind-FlowSGS against existing blind posterior samplers in~\Cref{tab:blind}.
We find that our method accurately estimates both the source galaxy image $\mathbf{x}$ and the lens mass parameters $\boldsymbol{\psi}$.

\vspace{-1.7mm}
\section{Discussion and Conclusion}
\label{sec:dis}
\vspace{-1.7mm}

In this work, we introduce Blind-FlowSGS, a principled posterior sampler for probabilistic inversion of strong gravitational lenses. 
With the incorporation of a flow matching prior trained from IllustrisTNG data, our method can sample from the posterior distribution more efficiently than existing diffusion-based methods without loss of accuracy.
For future work, we plan to apply our method to real lenses in the Euclid Q1 data release~\cite{Euclid_engineA} and compare our modeling results with the provided catalog, as our current test sources come from the same simulation pipeline that trains our prior and our test lens parameters are drawn from the same prior used at inference, so the present results represent a best case. Moving to more expressive parameterizations of the lens mass, such as an elliptical power law (EPL) with external shear or subhalo substructure, is also a natural next step.
Another direction is to train the flow matching prior on realistic galaxy images, which might suffer from non-ideal effects like noise and PSF blurring~\cite{li2023galaxy} but provide a more accurate prior than simulated data. 
On the methodology side, future directions include learning a lens-parameter regressor~\cite{euclid2026lemon} for initialization, accounting for unknown source amplitude and lens light effects, and increasing scalability with more efficient flow network architectures~\cite{li2026shuffleflow}.
Our metrics measure reconstruction accuracy rather than posterior correctness, so we also plan to evaluate Blind-FlowSGS with uncertainty calibration tests such as TARP~\cite{lemos2023sampling} and study theoretical guarantees.

\begin{ack}
We gratefully acknowledge the support of the NSF-Simons AI Institute for the Sky (SkAI) via grants NSF AST-2421845 and Simons Foundation MPS-AI-00010513.
This material is based upon work supported by the U.S. National Science Foundation under Award No. 2542022.
This research used the Delta advanced computing and data resource, which is supported by the National Science Foundation (award OAC 2005572) and the State of Illinois. Delta is a joint effort of the University of Illinois Urbana-Champaign and its National Center for Supercomputing Applications.
This work was produced by Fermi Forward Discovery Group, LLC under Contract No. 89243024CSC000002 with the U.S. Department of Energy, Office of Science, Office of High Energy Physics. Publisher acknowledges the U.S. Government license to provide public access under the DOE Public Access Plan
(\url{http://energy.gov/downloads/doe-public-access-plan}).
The authors would like to thank Yu Sun for helpful discussions and 4 anonymous reviewers for helpful feedback.
\end{ack}

{
    \small
    \bibliographystyle{plain}
    \bibliography{reference.bib}

\begin{thebibliography}{10}

\bibitem{adam2022posterior}
Alexandre {Adam}, Adam {Coogan}, Nikolay {Malkin}, Ronan {Legin}, Laurence {Perreault-Levasseur}, Yashar {Hezaveh}, and Yoshua {Bengio}.
\newblock {Posterior samples of source galaxies in strong gravitational lenses with score-based priors}.
\newblock In {\em Machine Learning and the Physical Sciences Workshop}, page arXiv.2211.03812, January 2022.

\bibitem{albergo2023building}
Michael~Samuel Albergo and Eric Vanden-Eijnden.
\newblock Building normalizing flows with stochastic interpolants.
\newblock In {\em The Eleventh International Conference on Learning Representations}, 2023.

\bibitem{baes2024tngskirtatlas}
Maarten {Baes}, Andrea {Gebek}, Ana {Tr{\v{c}}ka}, Peter {Camps}, Arjen {van der Wel}, {Abdurro'uf}, Nick {Andreadis}, Sena~Bokona {Tulu}, Abdissa~Tassama {Emana}, Jacopo {Fritz}, Raymond {Kelly}, Inja {Kova{\v{c}}i{\'c}}, Antonio {La Marca}, Marco {Martorano}, Aleksandr {Mosenkov}, Angelos {Nersesian}, Vicente {Rodriguez-Gomez}, Crescenzo {Tortora}, Bert {Vander Meulen}, and Lingyu {Wang}.
\newblock {The TNG50-SKIRT Atlas: Post-processing methodology and first data release}.
\newblock {\em \aap}, 683:A181, March 2024.

\bibitem{barco2025blind}
Gabriel~Missael Barco, Ronan Legin, Connor Stone, Yashar Hezaveh, and Laurence Perreault-Levasseur.
\newblock Blind strong gravitational lensing inversion: Joint inference of source and lens mass with score-based models.
\newblock {\em arXiv preprint arXiv:2511.04792}, 2025.

\bibitem{boyd2004convex}
Stephen Boyd and Lieven Vandenberghe.
\newblock {\em Convex optimization}.
\newblock Cambridge university press, 2004.

\bibitem{brewer2006strong}
Brendon~J Brewer and Geraint~F Lewis.
\newblock {Strong gravitational lens inversion: a Bayesian approach}.
\newblock {\em The Astrophysical Journal}, 637(2):608--619, 2006.

\bibitem{chan2016plug}
Stanley~H Chan, Xiran Wang, and Omar~A Elgendy.
\newblock {Plug-and-play ADMM for image restoration: Fixed-point convergence and applications}.
\newblock {\em IEEE Transactions on Computational Imaging}, 3(1):84--98, 2016.

\bibitem{charbonnier1994two}
Pierre Charbonnier, Laure Blanc-Feraud, Gilles Aubert, and Michel Barlaud.
\newblock Two deterministic half-quadratic regularization algorithms for computed imaging.
\newblock In {\em Proceedings of 1st international conference on image processing}, volume~2, pages 168--172. IEEE, 1994.

\bibitem{chung2023diffusion}
Hyungjin Chung, Jeongsol Kim, Michael~Thompson Mccann, Marc~Louis Klasky, and Jong~Chul Ye.
\newblock Diffusion posterior sampling for general noisy inverse problems.
\newblock In {\em The Eleventh International Conference on Learning Representations}, 2023.

\bibitem{Collett_2015}
Thomas~E. Collett.
\newblock The population of galaxy–galaxy strong lenses in forthcoming optical imaging surveys.
\newblock {\em The Astrophysical Journal}, 811(1):20, September 2015.

\bibitem{Einstein1936LENSLIKEAO}
Albert~B. Einstein.
\newblock Lens-like action of a star by the deviation of light in the gravitational field.
\newblock {\em Science}, 84 2188:506--7, 1936.

\bibitem{Erickson_2025}
Sydney Erickson, Sebastian Wagner-Carena, Phil Marshall, Martin Millon, Simon Birrer, Aaron Roodman, Thomas Schmidt, Tommaso Treu, Stefan Schuldt, Anowar~J. Shajib, and Padma~and Venkatraman.
\newblock Lens modeling of strides strongly lensed quasars using neural posterior estimation.
\newblock {\em The Astronomical Journal}, 170(1):44, June 2025.

\bibitem{euclid2026lemon}
{Euclid Collaboration}.
\newblock {Euclid Quick Data Release (Q1) XXXI. LEMON--LEns MOdelling with Neural networks. Automated and fast modelling of Euclid gravitational lenses with singular isothermal ellipsoid mass profile}.
\newblock {\em Astronomy \& Astrophysics}, 711:A31, 2026.

\bibitem{Euclid_engineA}
{Euclid Collaboration}, {Walmsley, M.}, {Holloway, P.}, {Lines, N. E. P.}, {Rojas, K.}, {Collett, T. E.}, {Verma, A.}, {Li, T.}, {Nightingale, J. W.}, {Despali, G.}, {Schuldt, S.}, {Gavazzi, R.}, {Melo, A.}, {Metcalf, R. B.}, {Andika, I. T.}, {Leuzzi, L.}, {Manjón-García, A.}, {Pearce-Casey, R.}, {Vincken, S. H.}, {Wilde, J.}, {Busillo, V.}, {Tortora, C.}, {Acevedo Barroso, J. A.}, {Dole, H.}, {Ecker, L. R.}, {Pearson, J.}, {Marshall, P. J.}, {More, A.}, {Saifollahi, T.}, {Gracia-Carpio, J.}, {Baeten, E.}, {Cornen, C.}, {Johnson, L. C.}, {Macmillan, C.}, {Kruk, S.}, {Remmelgas, K. A.}, {Clément, B.}, {Degaudenzi, H.}, {Courbin, F.}, {Bovy, J.}, {Casas, S.}, {Dannerbauer, H.}, {Diego, J. M.}, {Finner, K.}, {Galan, A.}, {Giocoli, C.}, {Hogg, N. B.}, {Jahnke, K.}, {Katona, J.}, {Kovács, A.}, {De Leo, C.}, {Mahler, G.}, {Millon, M.}, {Nagam, B. C.}, {Nugent, P.}, {Sainz de Murieta, A.}, {O’Riordan, C. M.}, {Sluse, D.}, {Sonnenfeld, A.}, {Spiniello, C.}, {Serjeant, S.}, {Thai, T. T.}, {Ulivi, L.}, {Walth, G. L.}, {Weisenbach, L.}, {Zumalacarregui, M.}, {Aghanim, N.}, {Altieri, B.}, {Amara, A.}, {Andreon, S.}, {Auricchio, N.}, {Aussel, H.}, {Baccigalupi, C.}, {Baldi, M.}, {Balestra, A.}, {Bardelli, S.}, {Battaglia, P.}, {Bernardeau, F.}, {Biviano, A.}, {Bonchi, A.}, {Bonino, D.}, {Branchini, E.}, {Brescia, M.}, {Brinchmann, J.}, {Camera, S.}, {Cañas-Herrera, G.}, {Capobianco, V.}, {Carbone, C.}, {Cardone, V. F.}, {Carretero, J.}, {Castander, F. J.}, {Castellano, M.}, {Castignani, G.}, {Cavuoti, S.}, {Chambers, K. C.}, {Cimatti, A.}, {Colodro-Conde, C.}, {Congedo, G.}, {Conselice, C. J.}, {Conversi, L.}, {Copin, Y.}, {Corcione, L.}, {Courtois, H. M.}, {Cropper, M.}, {Da Silva, A.}, {De Lucia, G.}, {Di Giorgio, A. M.}, {Dolding, C.}, {Dubath, F.}, {Duncan, C. A. J.}, {Dupac, X.}, {Ealet, A.}, {Escoffier, S.}, {Fabricius, M.}, {Farina, M.}, {Farinelli, R.}, {Faustini, F.}, {Finelli, F.}, {Fotopoulou, S.}, {Frailis, M.}, {Franceschi, E.}, {Fumana, M.}, {Galeotta, S.}, {George, K.}, {Gillard, W.}, {Gillis, B.}, {Gómez-Alvarez, P.}, {Granett, B. R.}, {Grazian, A.}, {Grupp, F.}, {Guzzo, L.}, {Gwyn, S.}, {Haugan, S. V. H.}, {Hoekstra, H.}, {Holmes, W.}, {Hook, I. M.}, {Hormuth, F.}, {Hornstrup, A.}, {Hudelot, P.}, {Jhabvala, M.}, {Joachimi, B.}, {Keihänen, E.}, {Kermiche, S.}, {Kiessling, A.}, {Kubik, B.}, {Kümmel, M.}, {Kunz, M.}, {Kurki-Suonio, H.}, {Lahav, O.}, {Le Boulc’h, Q.}, {Le Brun, A. M. C.}, {Le Mignant, D.}, {Ligori, S.}, {Lilje, P. B.}, {Lindholm, V.}, {Lloro, I.}, {Mainetti, G.}, {Maino, D.}, {Maiorano, E.}, {Mansutti, O.}, {Marcin, S.}, {Marggraf, O.}, {Martinelli, M.}, {Martinet, N.}, {Marulli, F.}, {Massey, R.}, {Maurogordato, S.}, {McCracken, H. J.}, {Medinaceli, E.}, {Mei, S.}, {Mellier, Y.}, {Meneghetti, M.}, {Merlin, E.}, {Meylan, G.}, {Mora, A.}, {Moresco, M.}, {Moscardini, L.}, {Nakajima, R.}, {Neissner, C.}, {Nichol, R. C.}, {Niemi, S.-M.}, {Padilla, C.}, {Paltani, S.}, {Pasian, F.}, {Pedersen, K.}, {Percival, W. J.}, {Pettorino, V.}, {Pires, S.}, {Polenta, G.}, {Poncet, M.}, {Popa, L. A.}, {Pozzetti, L.}, {Raison, F.}, {Rebolo, R.}, {Renzi, A.}, {Rhodes, J.}, {Riccio, G.}, {Romelli, E.}, {Roncarelli, M.}, {Saglia, R.}, {Sakr, Z.}, {Sánchez, A. G.}, {Sapone, D.}, {Sartoris, B.}, {Schewtschenko, J. A.}, {Schirmer, M.}, {Schneider, P.}, {Schrabback, T.}, {Secroun, A.}, {Seidel, G.}, {Seiffert, M.}, {Serrano, S.}, {Simon, P.}, {Sirignano, C.}, {Sirri, G.}, {Spurio Mancini, A.}, {Stanco, L.}, {Steinwagner, J.}, {Tallada-Crespí, P.}, {Taylor, A. N.}, {Tereno, I.}, {Tessore, N.}, {Toft, S.}, {Toledo-Moreo, R.}, {Torradeflot, F.}, {Tutusaus, I.}, {Valentijn, E. A.}, {Valenziano, L.}, {Valiviita, J.}, {Vassallo, T.}, {Verdoes Kleijn, G.}, {Veropalumbo, A.}, {Wang, Y.}, {Weller, J.}, {Zacchei, A.}, {Zamorani, G.}, {Zerbi, F. M.}, {Zucca, E.}, {Allevato, V.}, {Ballardini, M.}, {Bolzonella, M.}, {Bozzo, E.}, {Burigana, C.}, {Cabanac, R.}, {Cappi, A.}, {Di Ferdinando, D.}, {Escartin Vigo, J. A.}, {Gabarra, L.}, {Huertas-Company, M.}, {Martín-Fleitas, J.}, {Matthew, S.}, {Mauri, N.}, {Pezzotta, A.}, {Pöntinen, M.}, {Porciani, C.}, {Risso, I.}, {Scottez, V.}, {Sereno, M.}, {Tenti, M.}, {Viel, M.}, {Wiesmann, M.}, {Akrami, Y.}, {Anselmi, S.}, {Archidiacono, M.}, {Atrio-Barandela, F.}, {Benoist, C.}, {Benson, K.}, {Bergamini, P.}, {Bertacca, D.}, {Bethermin, M.}, {Blot, L.}, {Borgani, S.}, {Brown, M. L.}, {Bruton, S.}, {Calabro, A.}, {Camacho Quevedo, B.}, {Caro, F.}, {Carvalho, C. S.}, {Castro, T.}, {Charles, Y.}, {Cogato, F.}, {Cooray, A. R.}, {Cucciati, O.}, {Davini, S.}, {De Paolis, F.}, {Desprez, G.}, {Díaz-Sánchez, A.}, {Diaz, J. J.}, {Di Domizio, S.}, {Enia, A.}, {Fang, Y.}, {Ferrari, A. G.}, {Finoguenov, A.}, {Fontana, A.}, {Franco, A.}, {Ganga, K.}, {García-Bellido, J.}, {Gasparetto, T.}, {Gautard, V.}, {Gaztanaga, E.}, {Giacomini, F.}, {Gozaliasl, G.}, {Guidi, M.}, {Gutierrez, C. M.}, {Hall, A.}, {Hartley, W. G.}, {Hemmati, S.}, {Hernández-Monteagudo, C.}, {Hildebrandt, H.}, {Hjorth, J.}, {Kajava, J. J. E.}, {Kang, Y.}, {Kansal, V.}, {Karagiannis, D.}, {Kiiveri, K.}, {Kirkpatrick, C. C.}, {Le Graet, J.}, {Legrand, L.}, {Lembo, M.}, {Lepori, F.}, {Leroy, G.}, {Lesci, G. F.}, {Lesgourgues, J.}, {Liaudat, T. I.}, {Loureiro, A.}, {Macias-Perez, J.}, {Maggio, G.}, {Magliocchetti, M.}, {Magnier, E. A.}, {Mannucci, F.}, {Maoli, R.}, {Martins, C. J. A. P.}, {Maurin, L.}, {Miluzio, M.}, {Monaco, P.}, {Moretti, C.}, {Morgante, G.}, {Murray, C.}, {Nadathur, S.}, {Naidoo, K.}, {Navarro-Alsina, A.}, {Nesseris, S.}, {Passalacqua, F.}, {Paterson, K.}, {Patrizii, L.}, {Pisani, A.}, {Potter, D.}, {Quai, S.}, {Radovich, M.}, {Rocci, P.-F.}, {Sacquegna, S.}, {Sahlén, M.}, {Sanders, D. B.}, {Sarpa, E.}, {Scarlata, C.}, {Schaye, J.}, {Schneider, A.}, {Sciotti, D.}, {Sellentin, E.}, {Shankar, F.}, {Smith, L. C.}, {Tanidis, K.}, {Testera, G.}, {Teyssier, R.}, {Tosi, S.}, {Troja, A.}, {Tucci, M.}, {Valieri, C.}, {Venhola, A.}, {Vergani, D.}, {Vernardos, G.}, {Verza, G.}, {Vielzeuf, P.}, {Walton, N. A.}, and {Scott, D.}
\newblock {Euclid Quick Data Release (Q1) - XXVI. The Strong Lensing Discovery Engine A – System overview and first lens sample}.
\newblock {\em \aap}, 711:A26, 2026.

\bibitem{geman1995nonlinear}
Donald Geman and Chengda Yang.
\newblock Nonlinear image recovery with half-quadratic regularization.
\newblock {\em IEEE transactions on Image Processing}, 4(7):932--946, 1995.

\bibitem{Hezaveh_2017}
Yashar~D. Hezaveh, Laurence~Perreault Levasseur, and Philip~J. Marshall.
\newblock Fast automated analysis of strong gravitational lenses with convolutional neural networks.
\newblock {\em Nature}, 548(7669):555–557, August 2017.

\bibitem{ho2020denoising}
Jonathan Ho, Ajay Jain, and Pieter Abbeel.
\newblock Denoising diffusion probabilistic models.
\newblock {\em Advances in neural information processing systems}, 33:6840--6851, 2020.

\bibitem{holzschuh2024flowmatching}
Benjamin Holzschuh and Nils Thuerey.
\newblock Flow matching for posterior inference with simulator feedback.
\newblock {\em arXiv preprint arXiv:2410.22573}, 2024.

\bibitem{hu2026prism}
Yuanyun Hu, Evan Bell, Guijin Wang, and Yu~Sun.
\newblock Prism: Probabilistic and robust inverse solver with measurement-conditioned diffusion prior for blind inverse problems.
\newblock In {\em ICASSP 2026-2026 IEEE International Conference on Acoustics, Speech and Signal Processing (ICASSP)}, pages 11432--11436. IEEE, 2026.

\bibitem{karchev2022strong}
Konstantin Karchev, Noemi~Anau Montel, Adam Coogan, and Christoph Weniger.
\newblock Strong-lensing source reconstruction with denoising diffusion restoration models.
\newblock {\em arXiv preprint arXiv:2211.04365}, 2022.

\bibitem{kim2025flowdps}
Jeongsol Kim, Bryan~Sangwoo Kim, and Jong~Chul Ye.
\newblock Flowdps: Flow-driven posterior sampling for inverse problems.
\newblock In {\em Proceedings of the IEEE/CVF International Conference on Computer Vision}, pages 12328--12337, 2025.

\bibitem{Legin_2021}
Ronan Legin, Yashar Hezaveh, Laurence~Perreault Levasseur, and Benjamin Wandelt.
\newblock {Simulation-Based Inference of Strong Gravitational Lensing Parameters}.
\newblock {\em arXiv preprint arXiv:2112.05278}, 2021.

\bibitem{lemos2023sampling}
Pablo Lemos, Adam Coogan, Yashar Hezaveh, and Laurence Perreault-Levasseur.
\newblock Sampling-based accuracy testing of posterior estimators for general inference.
\newblock In {\em International Conference on Machine Learning}, pages 19256--19273. PMLR, 2023.

\bibitem{Li_2024_largepopulation}
Tian Li, Thomas~E Collett, Coleman~M Krawczyk, and Wolfgang Enzi.
\newblock Cosmology from large populations of galaxy–galaxy strong gravitational lenses.
\newblock {\em Monthly Notices of the Royal Astronomical Society}, 527(3):5311--5323, January 2024.

\bibitem{li2023galaxy}
Tianao Li and Emma Alexander.
\newblock {Galaxy image deconvolution for weak gravitational lensing with unrolled plug-and-play ADMM}.
\newblock {\em Monthly Notices of the Royal Astronomical Society: Letters}, 522(1):L31--L35, 2023.

\bibitem{li2026flowsgs}
Tianao Li, Xinhui Qian, and Emma Alexander.
\newblock Flowsgs: Improving flow priors for inverse imaging with stochastic interpolants.
\newblock {\em arXiv preprint arXiv:2609.20769}, 2026.

\bibitem{li2026shuffleflow}
Tianao Li, Tjitske Starkenburg, Yu~Sun, and Emma Alexander.
\newblock {ShuffleFlow: Scalable posterior inference for Bayesian inverse imaging}.
\newblock In {\em IEEE International Conference on Computational Photography (ICCP)}. IEEE, 2026.

\bibitem{lipman2023flow}
Yaron Lipman, Ricky T.~Q. Chen, Heli Ben-Hamu, Maximilian Nickel, and Matthew Le.
\newblock Flow matching for generative modeling.
\newblock In {\em The Eleventh International Conference on Learning Representations}, 2023.

\bibitem{liu2023flow}
Xingchao Liu, Chengyue Gong, and qiang liu.
\newblock Flow straight and fast: Learning to generate and transfer data with rectified flow.
\newblock In {\em The Eleventh International Conference on Learning Representations}, 2023.

\bibitem{martin2025pnpflow}
S{\'e}gol{\`e}ne~Tiffany Martin, Anne Gagneux, Paul Hagemann, and Gabriele Steidl.
\newblock Pnp-flow: Plug-and-play image restoration with flow matching.
\newblock In {\em The Thirteenth International Conference on Learning Representations}, 2025.

\bibitem{GravitationalLensing}
Georges {Meylan}, Philippe {Jetzer}, Pierre {North}, Peter {Schneider}, Christopher~S. {Kochanek}, and Joachim {Wambsganss}, editors.
\newblock {\em {Gravitational Lensing: Strong, Weak and Micro}}, January 2006.

\bibitem{murata2023gibbsddrm}
Naoki Murata, Koichi Saito, Chieh-Hsin Lai, Yuhta Takida, Toshimitsu Uesaka, Yuki Mitsufuji, and Stefano Ermon.
\newblock {GibbsDDRM: A partially collapsed Gibbs sampler for solving blind inverse problems with denoising diffusion restoration}.
\newblock In {\em International conference on machine learning}, pages 25501--25522. PMLR, 2023.

\bibitem{neal2011distributed}
Parikh Neal, Chu Eric, Peleato Borja, and Eckstein Jonathan.
\newblock Distributed optimization and statistical learning via the alternating direction method of multipliers.
\newblock {\em Foundations and Trends{\textregistered} in Machine learning}, 3(1):1--122, 2011.

\bibitem{nelson2019first}
Dylan Nelson, Annalisa Pillepich, Volker Springel, R{\"u}diger Pakmor, Rainer Weinberger, Shy Genel, Paul Torrey, Mark Vogelsberger, Federico Marinacci, and Lars Hernquist.
\newblock {First results from the TNG50 simulation: galactic outflows driven by supernovae and black hole feedback}.
\newblock {\em Monthly Notices of the Royal Astronomical Society}, 490(3):3234--3261, 2019.

\bibitem{nelson2021illustristngsimulationspublicdata}
Dylan {Nelson}, Volker {Springel}, Annalisa {Pillepich}, Vicente {Rodriguez-Gomez}, Paul {Torrey}, Shy {Genel}, Mark {Vogelsberger}, Ruediger {Pakmor}, Federico {Marinacci}, Rainer {Weinberger}, Luke {Kelley}, Mark {Lovell}, Benedikt {Diemer}, and Lars {Hernquist}.
\newblock {The IllustrisTNG simulations: public data release}.
\newblock {\em Computational Astrophysics and Cosmology}, 6(1):2, May 2019.

\bibitem{patel2025flowchef}
Maitreya Patel, Song Wen, Dimitris~N Metaxas, and Yezhou Yang.
\newblock Flowchef: Steering of rectified flow models for controlled generations.
\newblock In {\em Proceedings of the IEEE/CVF International Conference on Computer Vision}, pages 15308--15318, 2025.

\bibitem{Pearson_2021}
James Pearson, Jacob Maresca, Nan Li, and Simon Dye.
\newblock {Strong lens modelling: comparing and combining Bayesian neural networks and parametric profile fitting}.
\newblock {\em Monthly Notices of the Royal Astronomical Society}, 505(3):4362–4382, May 2021.

\bibitem{pillepich2019first}
Annalisa Pillepich, Dylan Nelson, Volker Springel, R{\"u}diger Pakmor, Paul Torrey, Rainer Weinberger, Mark Vogelsberger, Federico Marinacci, Shy Genel, Arjen van~der Wel, et~al.
\newblock {First results from the TNG50 simulation: the evolution of stellar and gaseous discs across cosmic time}.
\newblock {\em Monthly Notices of the Royal Astronomical Society}, 490(3):3196--3233, 2019.

\bibitem{Poh_2025}
Jason Poh, Ashwin Samudre, Aleksandra Ćiprijanović, Joshua Frieman, Gourav Khullar, and Brian~D. Nord.
\newblock Deep inference of simulated strong lenses in ground-based surveys.
\newblock {\em Journal of Cosmology and Astroparticle Physics}, 2025(05):053, May 2025.

\bibitem{potaptchik2026meta}
Peter Potaptchik, Adhi Saravanan, Abbas Mammadov, Alvaro Prat, Michael~Samuel Albergo, and Yee~Whye Teh.
\newblock Meta flow maps enable scalable reward alignment.
\newblock In {\em Forty-third International Conference on Machine Learning}, 2026.

\bibitem{roberts1996exponential}
Gareth~O. Roberts and Richard~L. Tweedie.
\newblock Exponential convergence of {L}angevin distributions and their discrete approximations.
\newblock {\em Bernoulli}, 2(4):341--363, 1996.

\bibitem{sanghvi2024kernel}
Yash Sanghvi, Yiheng Chi, and Stanley~H Chan.
\newblock Kernel diffusion: An alternate approach to blind deconvolution.
\newblock In {\em European Conference on Computer Vision}, pages 1--20. Springer, 2024.

\bibitem{sayed2025flowlensing}
Hamees Sayed, Pranath Reddy, Michael Toomey, and Sergei Gleyzer.
\newblock Flowlensing: Simulating gravitational lensing with flow matching.
\newblock {\em arXiv preprint arXiv:2510.07878}, 2025.

\bibitem{Schuldt_2023}
S.~Schuldt, R.~Cañameras, Y.~Shu, S.~H. Suyu, S.~Taubenberger, T.~Meinhardt, and L.~Leal-Taixé.
\newblock Holismokes: Ix. neural network inference of strong-lens parameters and uncertainties from ground-based images.
\newblock {\em Astronomy \& Astrophysics}, 671:A147, March 2023.

\bibitem{Shajib_2020}
A~J Shajib, S~Birrer, T~Treu, A~Agnello, E~J Buckley-Geer, J~H~H Chan, L~Christensen, C~Lemon, H~Lin, M~Millon, J~Poh, C~E Rusu, D~Sluse, C~Spiniello, G~C-F Chen, T~Collett, F~Courbin, C~D Fassnacht, J~Frieman, A~Galan, D~Gilman, A~More, T~Anguita, M~W Auger, V~Bonvin, R~McMahon, G~Meylan, K~C Wong, T~M~C Abbott, J~Annis, S~Avila, K~Bechtol, D~Brooks, D~Brout, D~L Burke, A~Carnero Rosell, M~Carrasco Kind, J~Carretero, F~J Castander, M~Costanzi, L~N da Costa, J~De Vicente, S~Desai, J~P Dietrich, P~Doel, A~Drlica-Wagner, A~E Evrard, D~A Finley, B~Flaugher, P~Fosalba, J~García-Bellido, D~W Gerdes, D~Gruen, R~A Gruendl, J~Gschwend, G~Gutierrez, D~L Hollowood, K~Honscheid, D~Huterer, D~J James, T~Jeltema, E~Krause, N~Kuropatkin, T~S Li, M~Lima, N~MacCrann, M~A~G Maia, J~L Marshall, P~Melchior, R~Miquel, R~L~C Ogando, A~Palmese, F~Paz-Chinchón, A~A Plazas, A~K Romer, A~Roodman, M~Sako, E~Sanchez, B~Santiago, V~Scarpine, M~Schubnell, D~Scolnic, S~Serrano, I~Sevilla-Noarbe, M~Smith, M~Soares-Santos, E~Suchyta, G~Tarle, D~Thomas, A~R Walker, and Y~Zhang.
\newblock {STRIDES: a 3.9 per cent measurement of the Hubble constant from the strong lens system DES J0408--5354}.
\newblock {\em Monthly Notices of the Royal Astronomical Society}, 494(4):6072–6102, March 2020.

\bibitem{Shajib_2024_slreview}
A.~J. Shajib, G.~Vernardos, T.~E. Collett, V.~Motta, D.~Sluse, L.~L.~R. Williams, P.~Saha, S.~Birrer, C.~Spiniello, and T.~Treu.
\newblock Strong lensing by galaxies.
\newblock {\em Space Science Reviews}, 220(8), November 2024.

\bibitem{Shajib_2023_SLfromVeraRubin}
Anowar~J. Shajib{\textdagger}, Graham~P. Smith, Simon Birrer, Aprajita Verma, Nikki Arendse, Thomas~E. Collett, Tansu Daylan, and Stephen Serjeant.
\newblock {Strong gravitational lenses from the Vera C. Rubin Observatory}.
\newblock {\em Phil. Trans. Roy. Soc. Lond. A}, 383(2295):20240117, 2025.

\bibitem{song2019generative}
Yang Song and Stefano Ermon.
\newblock Generative modeling by estimating gradients of the data distribution.
\newblock {\em Advances in neural information processing systems}, 32, 2019.

\bibitem{song2021scorebased}
Yang Song, Jascha Sohl-Dickstein, Diederik~P Kingma, Abhishek Kumar, Stefano Ermon, and Ben Poole.
\newblock Score-based generative modeling through stochastic differential equations.
\newblock In {\em International Conference on Learning Representations}, 2021.

\bibitem{stone2024caustics}
Connor {Stone}, Alexandre {Adam}, Adam {Coogan}, M.~{Yantovski-Barth}, Andreas {Filipp}, Landung {Setiawan}, Cordero {Core}, Ronan {Legin}, Charles {Wilson}, Gabriel {Barco}, Yashar {Hezaveh}, and Laurence {Perreault-Levasseur}.
\newblock {Caustics: A Python Package for Accelerated Strong Gravitational Lensing Simulations}.
\newblock {\em The Journal of Open Source Software}, 9(103):7081, November 2024.

\bibitem{treu2010strong}
Tommaso Treu.
\newblock Strong lensing by galaxies.
\newblock {\em Annual Review of Astronomy and Astrophysics}, 48:87--125, 2010.

\bibitem{Treu_2022}
Tommaso Treu, Sherry~H. Suyu, and Philip~J. Marshall.
\newblock Strong lensing time-delay cosmography in the 2020s.
\newblock {\em The Astronomy and Astrophysics Review}, 30(1), November 2022.

\bibitem{vegetti2023_sl_dm}
S.~Vegetti, S.~Birrer, G.~Despali, C.~D. Fassnacht, D.~Gilman, Y.~Hezaveh, L.~Perreault Levasseur, J.~P. McKean, D.~M. Powell, C.~M. O'Riordan, and G.~Vernardos.
\newblock Strong gravitational lensing as a probe of dark matter, 2023.

\bibitem{venkatakrishnan2013plug}
Singanallur~V Venkatakrishnan, Charles~A Bouman, and Brendt Wohlberg.
\newblock Plug-and-play priors for model based reconstruction.
\newblock In {\em 2013 IEEE global conference on signal and information processing}, pages 945--948. IEEE, 2013.

\bibitem{vono2019split}
Maxime Vono, Nicolas Dobigeon, and Pierre Chainais.
\newblock {Split-and-augmented Gibbs sampler—Application to large-scale inference problems}.
\newblock {\em IEEE Transactions on Signal Processing}, 67(6):1648--1661, 2019.

\bibitem{wu2024principled}
Zihui Wu, Yu~Sun, Yifan Chen, Bingliang Zhang, Yisong Yue, and Katherine~L Bouman.
\newblock Principled probabilistic imaging using diffusion models as plug-and-play priors.
\newblock {\em Advances in Neural Information Processing Systems}, 37:118389--118427, 2024.

\bibitem{zhang2025improving}
Bingliang Zhang, Wenda Chu, Julius Berner, Chenlin Meng, Anima Anandkumar, and Yang Song.
\newblock Improving diffusion inverse problem solving with decoupled noise annealing.
\newblock In {\em Proceedings of the Computer Vision and Pattern Recognition Conference}, pages 20895--20905, 2025.

\end{thebibliography}
}

\appendix

\newpage

\section{Lens parameter priors}
\label{app:prior}

\begin{figure}[h]
    \centering
    \includegraphics[width=\linewidth]{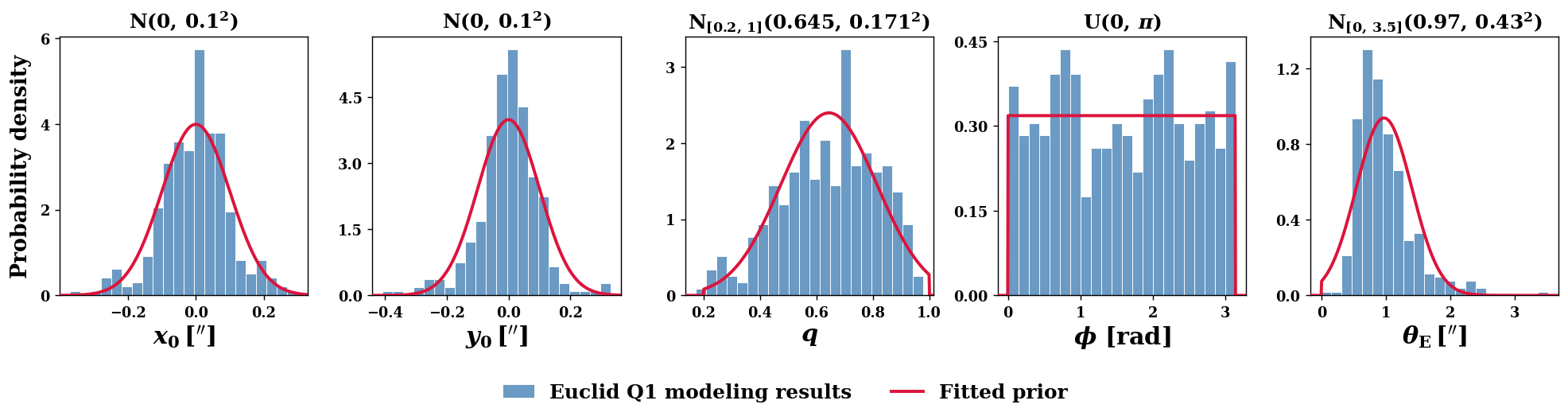}
    \caption{
        \textbf{SIE lens mass parameters estimated from Euclid Q1 modeling results~\cite{Euclid_engineA}.}
        We show both the histograms from the Euclid modeling results and the fitted prior distributions.
    }
    \label{fig:prior}
\end{figure}

We model the deflector mass as a singular isothermal ellipsoid (SIE, 5 parameters) and estimate prior distributions for these 5 parameters $\boldsymbol{\psi}=\{x_0, y_0, q, \phi,\theta_{\rm E}\}$ from the modeling results in the Euclid Q1 data release~\cite{Euclid_engineA} (see~\Cref{tab:priors} and~\Cref{fig:prior}).
We use a uniform distribution, two normal distributions, and two truncated normal distributions depending on the properties of different parameters.

\paragraph{Whitening map and projection.}
The parameter block of~\Cref{alg:blind-flowsgs} operates on whitened coordinates $\mathbf{u}=\mathcal{T}(\boldsymbol{\psi})$ rather than on $\boldsymbol{\psi}$ directly.
We take $\mathcal{T}$ to be the coordinate-wise affine standardization of the priors in~\Cref{tab:priors},
\begin{equation}
    u_i = \mathcal{T}_i(\psi_i) = \frac{\psi_i - m_i}{s_i},
    \qquad
    \psi_i = \mathcal{T}_i^{-1}(u_i) = m_i + s_i\,u_i,
    \label{eq:whitening}
\end{equation}
where $(m_i, s_i)$ are the mean and standard deviation of the (untruncated) Gaussian prior for $x_0$, $y_0$, $q$, and $\theta_{\rm E}$, and the midpoint and half-width of the support for the uniform prior on $\phi$, i.e.\ $m_\phi = s_\phi = \pi/2$.
In these coordinates the centroid priors become $\mathcal{N}(0,1)$ (e.g.\ $u_i=\psi_i/0.1$), the priors on $q$ and $\theta_{\rm E}$ become standard normals truncated to the whitened support $[\ell_i, h_i] = [(a_i - m_i)/s_i,\,(b_i - m_i)/s_i]$, and the prior on $\phi$ becomes $\mathcal{U}[-1,1]$.
Every coordinate therefore carries a prior scale of order one, so a single Langevin step size $\eta$ serves all 5 parameters even though their prior scales differ by about an order of magnitude, from $0.1$ for the centroid to $\pi/2$ for $\phi$.
Because $\mathcal{T}$ is affine and coordinate-wise, its Jacobian is constant, $\mathbf{J} = \partial\boldsymbol{\psi}/\partial\mathbf{u} = \mathrm{diag}(s_1,\dots,s_5)$, and likelihood gradients are pulled back as $\nabla_{\mathbf{u}} = \mathbf{J}\,\nabla_{\boldsymbol{\psi}}$.
The Langevin step in~\Cref{alg:blind-flowsgs} uses the exact prior score in whitened coordinates: $\nabla_{\mathbf{u}}\log p(\mathbf{u}) = -\mathbf{u}$ for the centroid and, within their truncated support, for $q$ and $\theta_{\rm E}$, since truncation changes only the normalizing constant, and $\mathbf{0}$ for $\phi$, whose whitened prior is uniform. Sampling the lens parameters directly rather than through an auxiliary split therefore avoids the Gaussian approximation that a conjugate prior step would require, at no extra cost, because the lens priors are cheap to evaluate in closed form.
The projection $\Pi$ maps the bounded coordinates $q$ and $\theta_{\rm E}$ back into $[\ell_i, h_i]$ by reflection at the boundaries, followed by a clamp to the same interval as a numerical safeguard. Because $\phi$ is $\pi$-periodic, $\Pi$ instead wraps its whitened coordinate into $[-1, 1)$, i.e.\ $\phi$ modulo $\pi$, so that position angles near $0$ and $\pi$ are treated as neighbors. The centroid coordinates are unbounded and left unchanged.

\begin{table}[h]
\centering
\caption{
    \textbf{Non-uniform priors on the SIE deflector mass parameters estimated from the modeling results of the Euclid Q1 data~\cite{Euclid_engineA}.}
}
\label{tab:priors}
\begin{tabular}{llll}
\toprule
\bf Parameter & \bf Symbol & \bf Prior & \bf Unit \\
\midrule
\multicolumn{4}{l}{\textit{Deflector mass (SIE)}} \\
Centroid        & $x_0$            & $\mathcal{N}(0,\,0.1^2)$                       & arcsec \\
                & $y_0$            & $\mathcal{N}(0,\,0.1^2)$                       & arcsec \\
Axis ratio      & $q$              & $\mathcal{N}_{[0.2,\,1.0]}(0.645,\,0.171^2)$    & \dots \\
Position angle  & $\phi$           & $\mathcal{U}(0,\,\pi)$                          & rad \\
Einstein radius & $\theta_{\rm E}$ & $\mathcal{N}_{[0,\,3.5]}(0.970,\,0.430^2)$      & arcsec \\
\bottomrule
\end{tabular}
\end{table}

\section{Implementation details}
\label{app:imp}

We show the pseudocode for FlowSGS in~\Cref{alg:flowsgs} and Blind-FlowSGS in~\Cref{alg:blind-flowsgs}.

\begin{algorithm}[h]
    \caption{Non-blind (source-only) posterior sampling with FlowSGS.}
    \label{alg:flowsgs}
    \begin{algorithmic}[1]
        \Require measurement $\mathbf{y}$, forward model $\mathcal{A}(\cdot)$, lens parameter $\boldsymbol{\psi}\in\mathbb{R}^{5}$, measurement noise level $\sigma$, velocity network $\mathbf{v}_\theta(\cdot,\cdot)$ pretrained with interpolant schedule $(\alpha_t, \sigma_t)$, diffusion coefficient $w_t$, number of iterations $K$, coupling parameters $\{\rho_k\}_{k=0}^{K-1}$
        \State $\mathbf{x}^0 \sim \mathcal{N}(0,\mathbf{I})$ \Comment{Initialization.}
        \For {$k = 0, 1, \ldots, K-1$}
            \State $\mathbf{z}^{k+1} \sim \pi(\mathbf{z} \vert \mathbf{x}^{k}) \propto \exp\left[-\frac{\|\mathcal{A}(\mathbf{z};\boldsymbol{\psi}) - \mathbf{y}\|^2}{2\sigma^2} - \frac{\|\mathbf{x}^{k} - \mathbf{z}\|^2}{2\rho_k^2}\right]$ \Comment{\textcolor{green}{\bf Likelihood step.}}
            \State \hspace{\algorithmicindent} $\mathcal{A}(\cdot\,;\boldsymbol{\psi})$ is affine in $\mathbf{z}$, so this Gaussian is sampled exactly by randomize-then-optimize.
            \State $t_k \leftarrow \left(\frac{\sigma_t}{\alpha_t}\right)^{-1}(\rho_k)$ \Comment{\textcolor{violet}{\bf Compute time $t_k$ corresponding to noise level $\rho_k$.}}
            \State $\mathbf{x}_{t_k} \leftarrow \alpha_{t_k}\, \mathbf{z}^{k+1}$ \Comment{\textcolor{violet}{\bf Scale $\mathbf{z}^{k+1}$ to match noise level at $t_k$.}}
            \State $\mathbf{x}^{k+1} \leftarrow \text{Euler–Maruyama}(\text{SDE},\mathbf{x}_{t_k},t_k,0)$ 
            \Comment{\textcolor{violet}{\bf Solve SDE with timestep correction.}}
            \State \hspace{\algorithmicindent} where the SDE is defined by 
            $\mathrm{d}\mathbf{x}_t 
            = \left[\left(1 - \frac{w_t \alpha_t}{2\sigma_t \gamma_t}\right)\mathbf{v}_\theta(\mathbf{x}_t, t) + \frac{w_t \dot\alpha_t}{2\sigma_t \gamma_t}\,\mathbf{x}_t\right]\mathrm{d}t + \sqrt{w_t}\,\mathrm{d}\bar{\mathbf{w}}_t,$
            \State \hspace{\algorithmicindent} and $\gamma_t = \dot\alpha_t\sigma_t - \alpha_t\dot\sigma_t$.
        \EndFor
        \State \Return $\mathbf{x}^K$
    \end{algorithmic}
\end{algorithm}

\section{Additional results}
\label{app:results}

\begin{figure*}[h]
    \centering
    \includegraphics[width=1.0\linewidth]{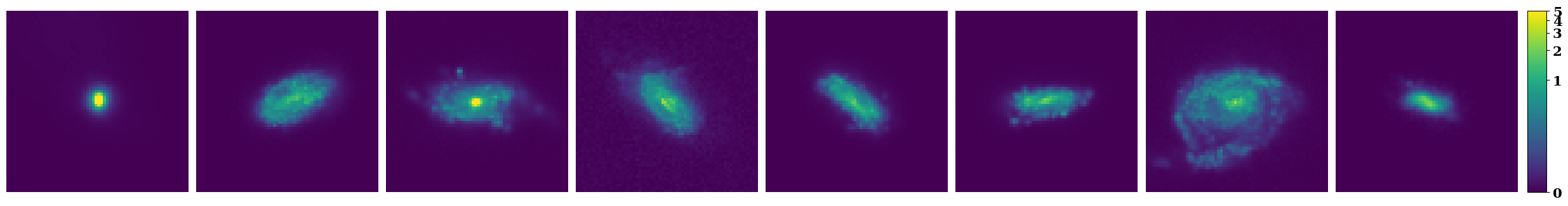}
    \caption{
        \textbf{Samples from the flow matching model trained on IllustrisTNG TNG50-1 data~\cite{nelson2019first, pillepich2019first, nelson2021illustristngsimulationspublicdata, baes2024tngskirtatlas}.}
    }
    \label{fig:samples}
\end{figure*}

\begin{figure*}[h]
    \centering
    \includegraphics[width=1.0\linewidth]{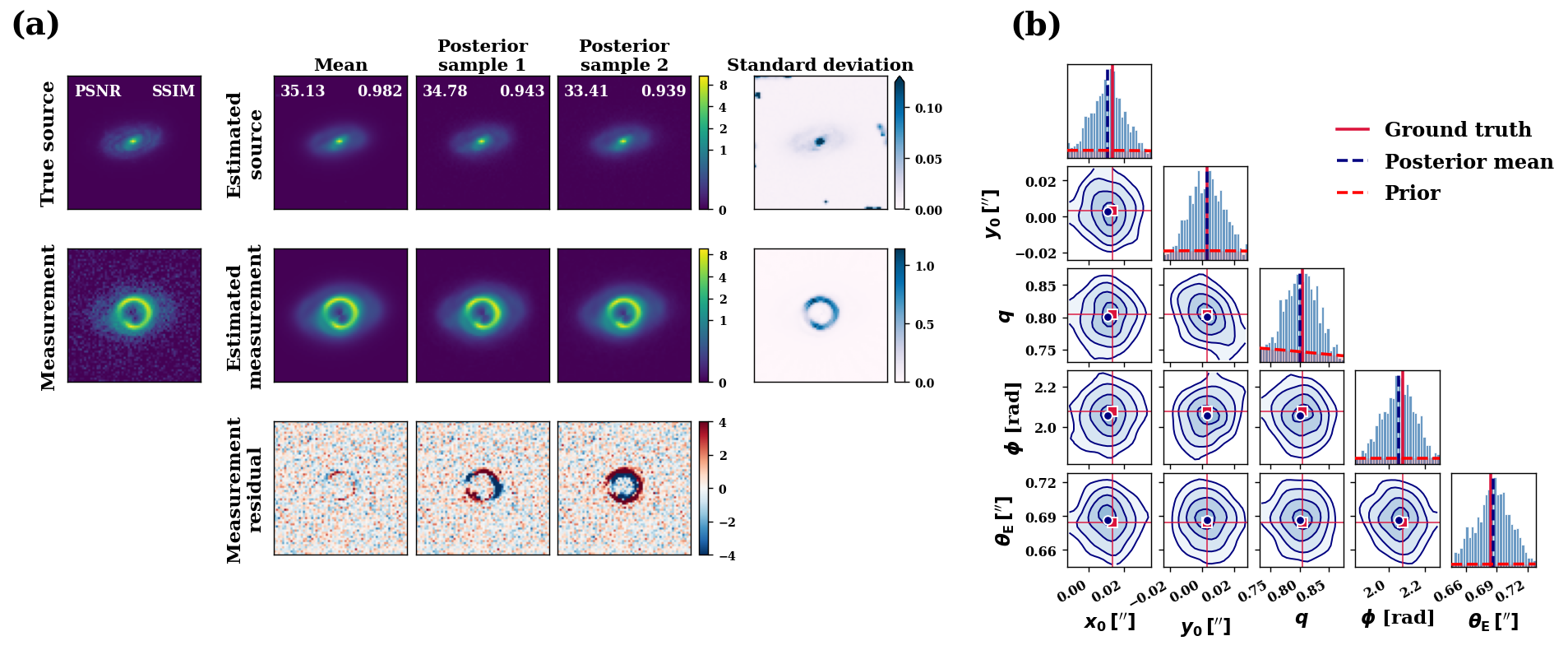}
    \includegraphics[width=1.0\linewidth]{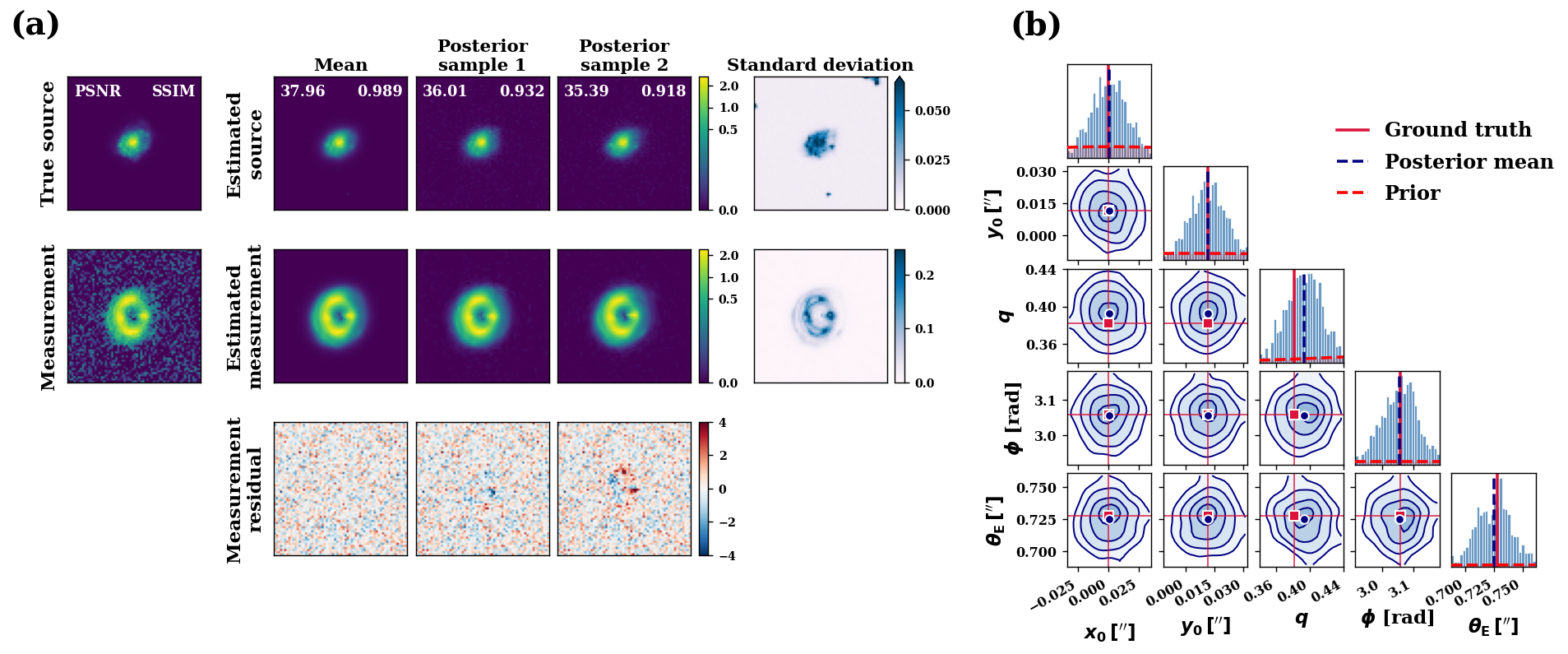}
    \includegraphics[width=1.0\linewidth]{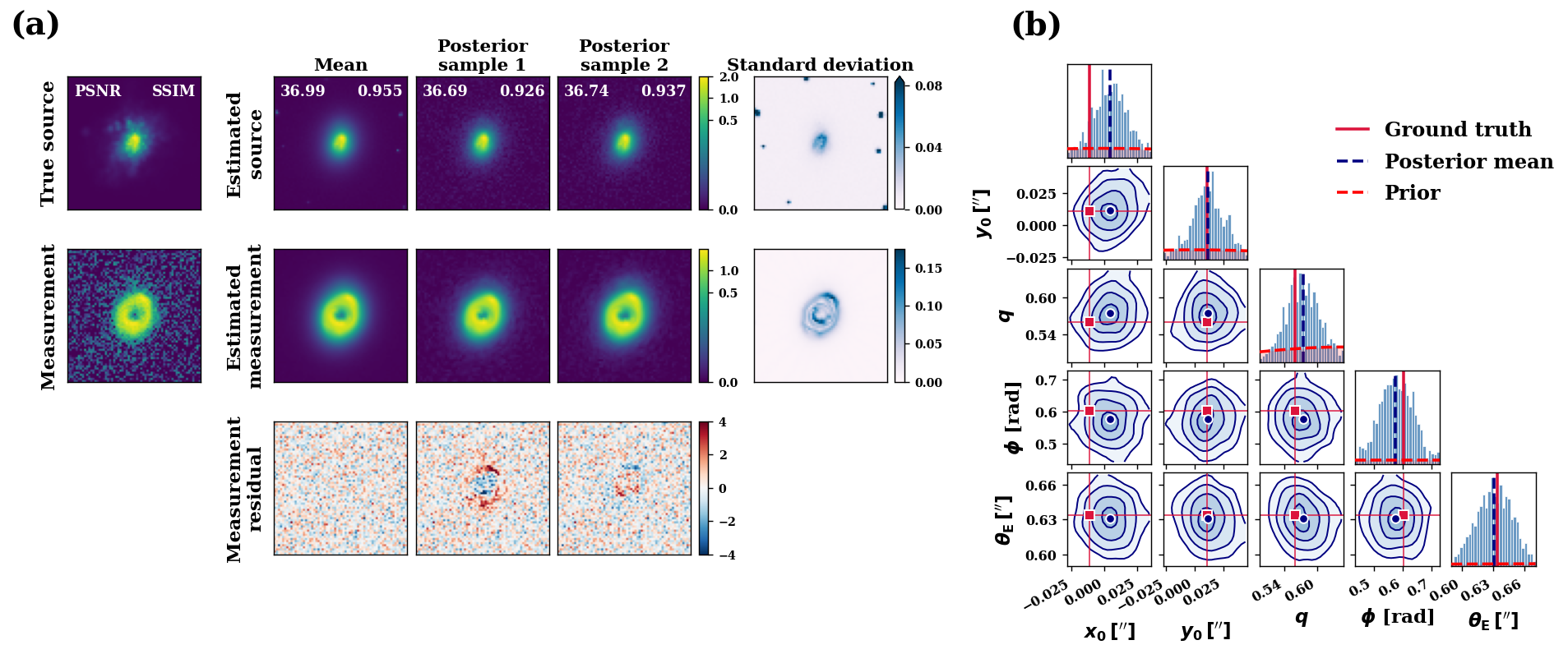}
    \caption{
        \textbf{Additional results on joint source and lens (blind) reconstruction with full corner plots.}
        (a) Source galaxy image, measurement, and measurement residuals, shown in mean, posterior samples, and standard deviations.
        (b) Corner plots for 5 SIE lens mass parameters computed from 1,024 posterior samples.
        Red dashed lines on the 1D marginals show the lens parameter priors of~\Cref{tab:priors}.
    }
    \label{fig:blind_supp}
\end{figure*}

We show unconditional samples from our pretrained flow matching prior in~\Cref{fig:samples}. We show additional results on joint reconstruction of the source $\mathbf{x}$ and the lens mass parameters $\boldsymbol{\psi}$ in~\Cref{fig:blind_supp}.

\end{document}